\documentclass[authoryear]{elsarticle}

\usepackage{hyperref}
\usepackage{lineno}
\modulolinenumbers[5]
\usepackage{bm}
\usepackage{amsmath}
\usepackage{subfigure}
\usepackage{color}
\usepackage{graphicx}
\usepackage{amssymb}
\usepackage{lscape}

\graphicspath{{./}}

\journal{Icarus}

\begin{document}

\begin{frontmatter}

\title{Giant Impact onto a Vesta-Like Asteroid and Formation of Mesosiderites through Mixing of Metallic Core and Surface Crust}

\author[mymainaddress]{Keisuke Sugiura\corref{mycorrespondingauthor}}
\cortext[mycorrespondingauthor]{Corresponding author}
\ead{sugiuraks@elsi.jp}

\author[myaddress2]{Makiko K. Haba}

\author[mymainaddress]{Hidenori Genda}

\address[mymainaddress]{Earth-Life Science Institute, Tokyo Institute of Technology, Tokyo 152-8550, Japan}

\address[myaddress2]{Department of Earth and Planetary Sciences, Tokyo Institute of Technology, Tokyo 152-8550, Japan}

\begin{abstract}
Mesosiderites are a type of stony-iron meteorites composed of a mixture of silicates and Fe-Ni metals. The mesosiderite silicates and metals are considered to have originated from the crust and metal core, respectively, of a differentiated asteroid. In contrast, mesosiderites rarely contain the olivine that is mainly included in a mantle. Although a giant impact onto a differentiated asteroid is considered to be a probable mechanism to mix crust and metal materials to form mesosiderites, it is not obvious how such a giant impact can form mesosiderite-like materials without including mantle materials. We conducted three-dimensional numerical simulations of giant impacts onto differentiated asteroids, using the smoothed particle hydrodynamics method, to investigate the detailed distribution of mixed materials on the resultant bodies. For the internal structure model of a target body, we used a thin-crust model derived from the magma ocean crystallization model of the asteroid Vesta. We also considered, as another possible internal structure for the target body, a thick crust and a large metal core suggested from the proximity observation of Vesta by the Dawn probe. In the simulations with the former model, excavation of the metal core requires nearly catastrophic impacts and mantle is exposed over large surface areas. Thus, stony-iron materials produced on its surface are likely to include mantle materials, and it is difficult to produce mesosiderite-like materials with this internal structure. Conversely, in the simulations with the latter model, mantle materials are exposed only at impact sites, even when the impacts excavate the metal core, and we confirmed that the formation of a surface with little mantle material and the formation of mesosiderite-like materials are possible from such a surface. Therefore, our simulations suggest that an internal structure with a thick crust and a large core is more likely as a mesosiderite parent body rather than the thin-crust internal structure inferred from the conventional magma ocean model.
\end{abstract}


\end{frontmatter}


\section{Introduction \label{Introduction}}
Mesosiderites are a type of stony-iron meteorite composed of silicates and Fe-Ni metals with a mean volume ratio of 7:3 (e.g.,\,\citealt{Mason-and-Jarosewich1973}). Because the silicate parts of mesosiderites display basaltic and gabbroic textures, the silicates are considered to have originated from the crust of a differentiated asteroid (e.g.,\,\citealt{Mittlefehldt-et-al1979}). The Fe-Ni metals of mesosiderites show chondritic siderophile element compositions with limited variability and well match to IIIAB iron meteorites, which suggests that they originated from the molten metallic core of a differentiated asteroid and were largely molten during mixing with the silicates (e.g.,\,\citealt{Hassanzadeh-et-al1990}). However, mesosiderites contain only small amounts of olivine, which is a main component of mantles, and the volume fraction of olivine of mesosiderites is at most 6\% (\citealt{Prinz-et-al1980}).

The mesosiderite silicates have basaltic and gabbroic textures with petrology and chemical compositions similar to those of howardite-eucrite-diogenite (HED) meteorites (e.g.,\,\citealt{Mittlefehldt-et-al1979}). Moreover, the O and Cr isotope compositions of mesosiderite silicates are in good agreement with those of HED meteorites (\citealt{Greenwood-et-al2006, Trinquier-et-al2007}). Thus, the parent body of mesosiderite silicates is the same as that of HED meteorites or another parent body with the same petrological and isotopic characteristic as the HED parent body. Because infrared spectral features of the asteroid Vesta are similar to those of HED meteorites, the parent body of HED meteorites is generally considered to be the asteroid Vesta (e.g.,\,\citealt{McCord-et-al1970, McSween-et-al2011}). Therefore, the parent body of mesosiderites is highly likely to be the asteroid Vesta as well. The high-precision U-Pb dating of mesosiderite zircons by isotope dilution thermal ionization mass spectrometry reveals that mesosiderite formation occurred $4{,}525.4 \pm 0.9\,{\rm Myr}$ ago, and the mesosiderite parent body should have a diameter larger than $500\,{\rm km}$ to have had a molten core at that time (\citealt{Haba-et-al2019}). The inferred diameter of the parent body is consistent with that of the asteroid Vesta. Note that the infrared spectral features of Vesta and HED meteorites are merely similar, and we cannot discard the possibility that HED meteorites, as well as mesosiderites, originated from another asteroid with similar spectral features (\citealt{Rubin-and-Mittlefehldt1993}).

A giant impact onto a differentiated asteroid is considered to be a promising mechanism for the formation of mesosiderites through mixing of the core and crust of the asteroid (e.g.,\,\citealt{Haack-et-al1996, Scott-et-al2001}). A parent body of mesosiderite metals may be different from that of silicates, but such a scenario is unlikely because the average impact velocity in the main asteroid belt $4{,}525\,{\rm Myr}$ ago was $\sim5\,{\rm km/s}$, which is much faster than the escape velocity from a $500\,{\rm km}$-sized asteroid, or $\sim300\,{\rm m/s}$. Note that the fraction of possible impacts with collisional velocities $< 300\,{\rm m/s}$ in the present main belt is only about 0.05\%, that is, such impacts are very rare (\citealt{Sugiura2020}, see also \citealt{Farinella-and-Davis1992, OBrien-and-Sykes2011}). A head-on impact may cause mixing of metals and silicates originating from different bodies, but a head-on collision between $500\,{\rm km}$-sized bodies with an impact velocity of $\sim5\,{\rm km/s}$ results in a catastrophic disruption, which significantly reduces sizes of colliding bodies. Very slow cooling rates of mesosiderites at low temperatures (0.4$^{\circ}$C/Myr at 400$^{\circ}$C, \citealt{Hopfe-and-Goldstein2001}) show that they were buried at depth in a $500\,{\rm km}$-sized asteroid after their formation (\citealt{Haack-et-al1996}). The smaller a parent body is, the faster cooling rate becomes (\citealt{Haack-et-al1990}), so that the giant impact that formed mesosiderites would not have significantly reduced the size of the parent body and thus catastrophic disruption of their parent body is an unlikely scenario. An oblique impact results in a hit-and-run collision, and thus the mixing of materials from different asteroids is typically very minimal. Although a chain of remnants produced through a hit-and-run collision may contain both metals and silicates originating from different bodies (see Fig.\,3b of \citealt{Asphaug-et-al2006}), such remnants are very small compared to parent bodies and are not appropriate for very slow cooling rate that mesosiderites experienced. Thus, although there are uncertainties in our discussion and other possibilities for mesosiderite formation scenarios, we argue that mesosiderite metals and silicates are likely to have come from the same parent body, and a possible formation scenario of mesosiderites is considered to be as follows (\citealt{Haba-et-al2019}): A giant impact onto a $500\,{\rm km}$-sized differentiated asteroid dug out its metal core, excavated metal materials reaccumulated on the original crust of the same asteroid, and the mixing of metal and crust materials formed mesosiderites. Our above discussion based on \cite{Haack-et-al1996} and \cite{Haba-et-al2019} implies that the giant impact is not likely to have appreciably reduced the size of the asteroid. Produced mesosiderite-like materials were ejected from those parent body through later cratering impacts and were delivered to the Earth as mesosiderite meteorites.

It is not obvious that such a mesosiderite formation scenario with keeping the given internal structure of the parent body is possible. Moreover, such a giant impact would also excavate mantle materials and produce stony-iron materials possibly containing considerable amounts of these mantle materials, which is not consistent with the composition of mesosiderites. We do not know whether it is possible that a giant impact mixes core materials and crust materials without including mantle materials. \cite{Scott-et-al2001} conducted low-resolution numerical simulations of giant impacts onto differentiated bodies using the smoothed particle hydrodynamics (SPH) method, but they did not investigate the spatial distribution of mixed materials in the impact outcomes.

In this study, we conducted three-dimensional numerical simulations of giant impacts onto differentiated bodies, and investigated the detailed distribution and composition of the produced mixed materials on the impact outcomes. Then we investigated whether it is possible that mesosiderite-like materials, that is, materials composed of metal and crust but not mantle materials, can be produced through a giant impact. Although there have been several studies investigating the distribution of materials after giant impacts onto differentiated bodies (e.g.,\,\citealt{Goblack-et-al2018, Emsenhuber-et-al2018}), we especially focus on the distribution of materials on the surfaces of resultant bodies after giant impacts and quantitatively investigate the fraction of materials on the surfaces.

\section{Method and initial conditions \label{Method-and-initial-conditions}}
\subsection{Method \label{Method}}
For an impact onto an asteroid with a diameter of about $500\,{\rm km}$, the outcome is dominated by the force of gravity rather than material strength, such as friction of granular materials or elastic forces of intact materials. Moreover, a large fraction of the metal core of a $500\,{\rm km}$-sized asteroid was still molten at $4{,}525\,{\rm Myr}$ ago (\citealt{Hassanzadeh-et-al1990, Haba-et-al2019}). Thus, we do not have to consider material strengths for the molten metal core. The crust and mantle of an asteroid were already solidified $4{,}525\,{\rm Myr}$ ago. Further, recent work shows that the material strengths surely affect collisional outcomes between $1{,}000\,{\rm km}$-sized asteroids, but the effect of the material strengths for distribution of materials within an impact outcome is limited (Fig.\,3 of \citealt{Emsenhuber-et-al2018}). Thus, we ignore the material strengths and consider collisions between fluid-like bodies without deviatoric stresses. Although the effect of the material strengths would become more important for our simulations with smaller targets, our code has not yet been fully tested for simulations with the mixture of fluid and solid materials, so that we leave giant impact simulations with the material strengths for our future work.

We used a classical SPH method for the fluid dynamics in our simulations (\citealt{Gingold-and-Monaghan1977, Lucy1977}). SPH methods are novel hydrodynamics simulation methods using particles that mimic parts of fluid bodies and are applied to impacts between planetary bodies (e.g.,\,\citealt{Genda-et-al2012, Carter-et-al2018}). The detailed equations we solved in our simulations are given in Appendix A. We used a leapfrog integrator as the time-advancing scheme (e.g.,\,\citealt{Hubber-et-al2013}). For a wide parameter survey using supercomputers, our simulation code was parallelized by exploiting the framework for developing particle simulator (FDPS: \citealt{Iwasawa-et-al2015, Iwasawa-et-al2016}).

To represent changes in the thermodynamic states of rocky or metallic materials during impacts, we used the Tillotson equation of state (\citealt{Tillotson1962}). A set of Tillotson parameters for iron (\citealt{Tillotson1962}) was applied to the SPH particles that compose the metallic cores. Although we distinguished SPH particles that compose surface crusts and mantle layers, the same set of Tillotson parameters for basalt (\citealt{Benz-and-Asphaug1999}) was applied to both crust and mantle SPH particles, because Tillotson parameters for different types of rocks (e.g.,\,basalt, granite, or peridotite) are similar and such minor differences in equations of state probably do not affect impact outcomes. The specific values of these Tillotson parameters are listed in Table \ref{tillotson-parameters}.

\begin{landscape}
\begin{table}[!htb]
  \begin{center}
    \begin{tabular}{c c c c c c c c c c c}\hline 
      & $\rho_{0}{\rm (g/cc)}$ & $u_{{\rm iv}}{\rm (erg/g)}$ & $u_{{\rm cv}}{\rm (erg/g)}$ & $u_{0}{\rm (erg/g)}$ & $A{\rm (erg/cc)}$ & $B{\rm (erg/cc)}$ & $a$ & $b$ & $\alpha$ & $\beta$ \rule[0mm]{0mm}{5mm}\\
      \hline
      core & $7.8$ & $2.4\times 10^{10}$ & $8.67\times 10^{10}$ & $9.5\times 10^{10}$ & $1.28\times 10^{12}$ & $1.05\times 10^{12}$ & $0.5$ & $1.5$ & $5.0$ & $5.0$ \rule[0mm]{0mm}{5mm}\\
      crust and mantle & $2.7$ & $4.72\times 10^{10}$ & $1.82\times 10^{11}$ & $4.87\times 10^{12}$ & $2.67\times 10^{11}$ & $2.67\times 10^{11}$ & $0.5$ & $1.5$ & $5.0$ & $5.0$ \rule[0mm]{0mm}{5mm}\\ 
      \hline
    \end{tabular}
  \end{center}
  \caption{Tillotson parameters applied in this study}
  \label{tillotson-parameters}
\end{table}
\end{landscape}

\subsection{Initial conditions \label{Initial-conditions}}
We used a differentiated asteroid with a diameter of about $500\,{\rm km}$ as a target body for impact simulations. A representative asteroid for such body is the asteroid Vesta, so we set the interior structure of the target body based on a magma-ocean crystallization model for Vesta with a chemical composition of HED meteorites (\citealt{Mandler-and-ElkinsTanton2013}). We set the radius of the target body, the radius of the metal core, the thickness of the mantle layer, and the thickness of the crust to be $270\,{\rm km}$, $110\,{\rm km}$, $120\,{\rm km}$, and $40\,{\rm km}$, respectively. We call impact simulations with this interior structure of the target body model 1.

However, the interior model is merely one possible configuration, and the actual interior structure depends on the initial composition of the building blocks and the differentiation process (\citealt{Ruzicka-et-al1997}). The proximity observation of Vesta by the Dawn probe suggests a different internal structure. Non-detection of olivine inside of the Rheasilvia crater and the subsequent simulation works suggest that the Vesta’s crust is thicker than $80\,{\rm km}$ (e.g.,\,\citealt{Clenet-et-al2014}, see also \citealt{Yamaguchi-et-al2011}). The radius of a metal core depends on its assumed density, and the gravitational moment $J_{2}$ of Vesta shows that its core radius is at most $140\,{\rm km}$, with the density of the core reduced due to sulfur\footnote{One could question whether compositions, especially FeS, of such a S-rich core are consistent with those of mesosiderites. However, the FeS contents in mesosiderites are not necessarily consistent with those of a whole S-rich core. The recent study suggests that planetesimal cores form a sulfide-rich immiscible liquid in molten core and thus sulfur may be heterogeneously distributed in a core (\citealt{Bercovici-et-al2019}). Because mesosiderite metals have sampled only small part of a core, their FeS contents probably do not match with those of a whole core.} (\citealt{Ermakov-et-al2014}). Thus, as another model for the formation of mesosiderites through a giant impact, we conducted numerical simulations with a target body having a core radius of $140\,{\rm km}$, a mantle thickness of $50\,{\rm km}$, and a crust thickness of $80\,{\rm km}$. We call these impact simulations model 2.

The central pressure of bodies with $250\,{\rm km}$ radii, $\sim100\,{\rm MPa}$, is much smaller than the bulk moduli of rocky or metallic materials $\sim10$--$100\,{\rm GPa}$. Thus, the density increase due to self-gravitational compression is negligible, and the central density of the bodies is similar to the uncompressed density. The initial densities of the metallic cores and the rocky layers of the target bodies were set to uncompressed densities calculated from the Tillotson equations of state for iron and basalt, respectively. We used about $1.1 \times 10^{5}$ SPH particles for the target body of model 1 and $1.3 \times 10^{5}$ SPH particles for the target body of model 2. Thus, the body was resolved by about 30 SPH particles along the radial direction, and the width of the surface crust layer was resolved by about 4 SPH particles for model 1 and 9 SPH particles for model 2. The radial resolution of the surface crust for model 1 is slightly coarse, but we confirmed that a higher resolution simulation for model 1 retains the same characteristics of a resultant body obtained from a simulation with the resolution here. Thus the slightly coarse resolution for the crust of model 1 is not a problem.

The effects of initial body rotation on the results of impacts are generally considered to be negligible. For example, Vesta's rotation period of $5.3\,{\rm h}$ induces a rotation speed at the surface of at most $100\,{\rm m/s}$, which is much smaller than the average impact velocity of $\sim5\,{\rm km/s}$ in the main belt, and the motion due to the rotation is negligible compared to the motion due to the impact itself. Thus, for simplicity, we did not give initial rotation to the target bodies.

We used a uniform basaltic sphere as an impactor. We investigated impacts with an impactor having a mass of $0.1 M_{{\rm target}}$, where $M_{{\rm target}}$ represents the target mass. We varied the impact velocity $v_{{\rm imp}}$ from $1\,{\rm km/s}$ to $5\,{\rm km/s}$, which covers typical impact velocities in the main belt. We varied the impact angle $\theta_{{\rm imp}}$ from $10^{\circ}$ to $50^{\circ}$, where $\theta_{{\rm imp}}=0^{\circ}$ represents a head-on collision and $\theta_{{\rm imp}}=90^{\circ}$ represents a fully grazing collision. Here, impact velocity vectors are parallel to the $x$-axis, and the centers of the initial targets and impactors are on the $xy$-plane. We conducted 85 runs of the impact simulations each for model 1 and model 2.

We continued each impact simulation until $5.0\times 10^{4}\,{\rm s}$ after the impact. Impacts deformed the target bodies and ejected fragments, and the timescale for relaxation of the configuration of deformed bodies and re-accumulation of fragments is roughly a free-fall timescale of $\sim10^{3}\,{\rm s}$. Thus, the simulation period of $5.0\times 10^{4}\,{\rm s}$ is sufficient to achieve relaxed configurations of the deformed bodies. We analyzed the properties of the body at $5.0\times 10^{4}\,{\rm s}$ after each impact.

\section{Results \label{Results}}

\subsection{Model 1: 110\,km core radius and 40\,km crust thickness\label{model1}}

\begin{figure}[!htb]
  \begin{center}
    \includegraphics[bb=0 0 1063 540, width=1.0\linewidth,clip]{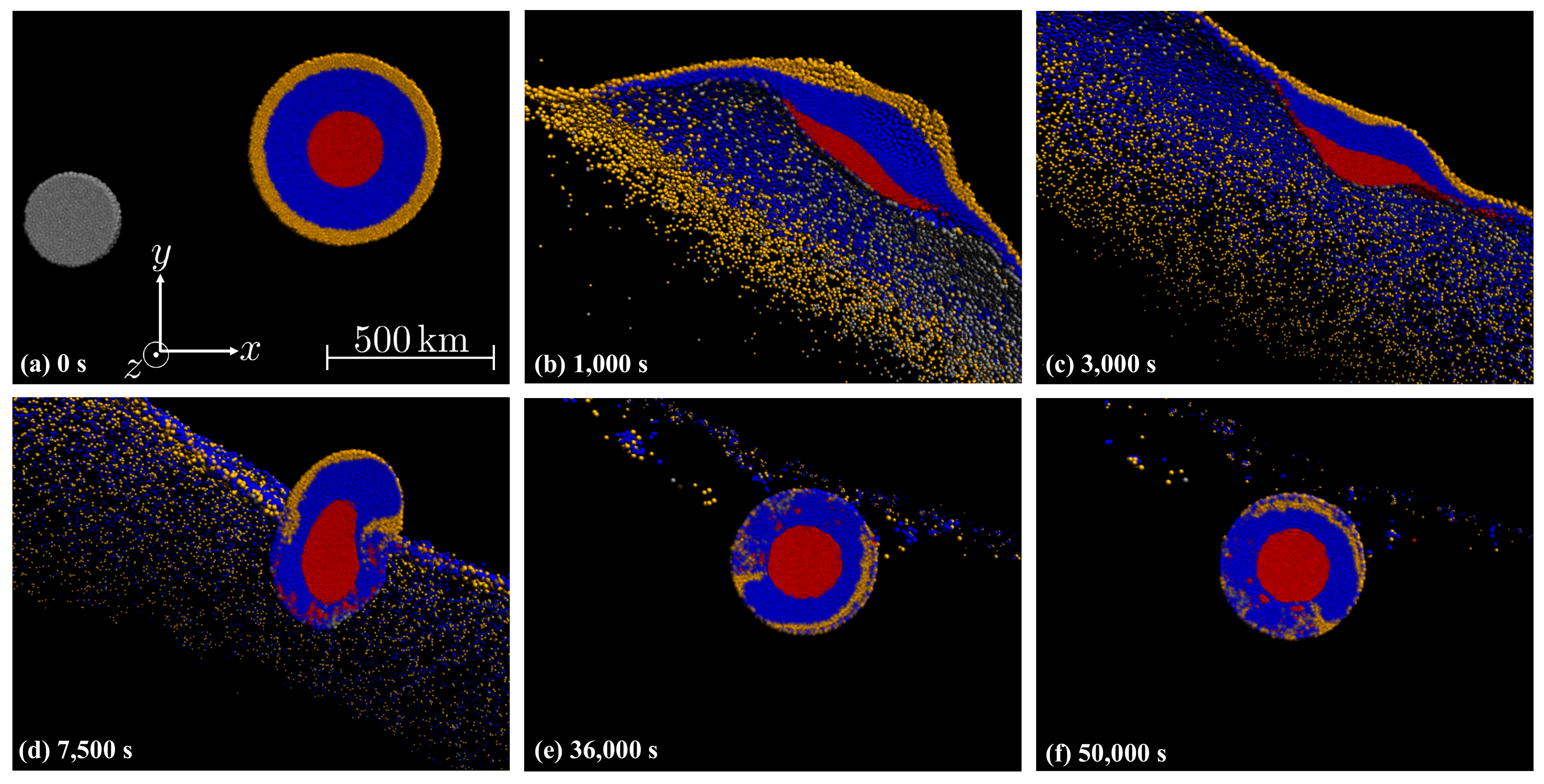}
    \caption{Snapshots of a simulation with model 1: giant impact with $\theta_{{\rm imp}}=30^{\circ}$ and $v_{{\rm imp}}=3.25\,{\rm km/s}$ onto a differentiated body with a core radius of 110\,km and crust thickness of 40\,km. Red, blue, yellow, and gray SPH particles represent the metallic core, the mantle layer, the surface crust, and the impactor, respectively. In panel (a), we show the coordinate system and the scale used in the simulation. We show cross sections of the impacting bodies; that is, only SPH particles behind the $xy$-plane are shown. Note that grey SPH particles are almost invisible after panel (b) because almost all of them are lost to space.}
    \label{subsequent-pictures-model1-30-3.25kms}
  \end{center}
\end{figure} 
Figure \ref{subsequent-pictures-model1-30-3.25kms} shows snapshots of an impact simulation of model 1, that is, a simulation with a differentiated target body with a 110\,km core radius and 40\,km crust thickness. The impact angle $\theta_{{\rm imp}}$ and impact velocity $v_{{\rm imp}}$ used in the simulation were $30^{\circ}$ and $3.25\,{\rm km/s}$, respectively. The target body was greatly deformed due to the impact (Fig.\,\ref{subsequent-pictures-model1-30-3.25kms}b), and the edges of the deformed metallic core were ejected from the target body (Fig.\,\ref{subsequent-pictures-model1-30-3.25kms}c). Some ejected metals reaccumulated on the original surface crust that survived the impact (Fig.\,\ref{subsequent-pictures-model1-30-3.25kms}d, e). The original mantle layer was exposed over almost half of the surface area of the resultant body (Fig.\,\ref{subsequent-pictures-model1-30-3.25kms}f). The resultant body had a mass of $\approx 0.5 M_{{\rm target}}$, and thus the impact resulted in nearly catastrophic disruption.

\begin{figure}[!htb]
  \begin{center}
    \includegraphics[bb=0 0 886 546, width=1.0\linewidth,clip]{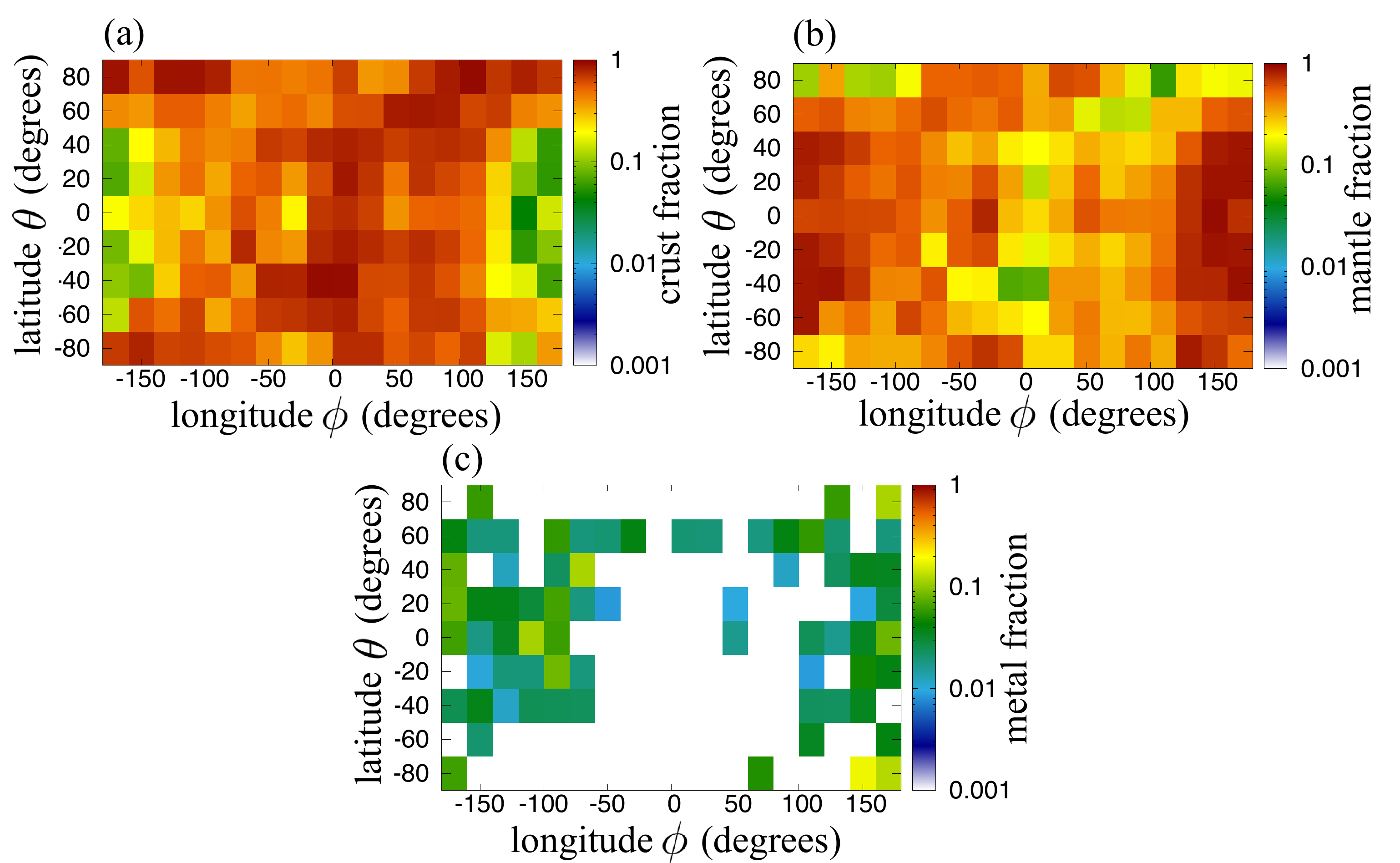}
    \caption{Mass fractions of materials on the surface of the resultant body for model 1: (a) crust, (b) mantle, and (c) metals. This surface was produced by a giant impact with $\theta_{{\rm imp}}=30^{\circ}$ and $v_{{\rm imp}}=3.25\,{\rm km/s}$ onto a differentiated body with a core radius of 110\,km and crust thickness of 40\,km. Here, the surface means materials at a depth of less than $20\,{\rm km}$. Latitudes $\theta$ (vertical axes) and longitudes $\phi$ (horizontal axes) show positions on the surface of the resultant body at $t=5.0\times 10^{4}\,{\rm s}$ (Fig.\,\ref{subsequent-pictures-model1-30-3.25kms}f). $\theta=0^{\circ}$ and $\phi=0^{\circ}$ mean $+x$-direction, $\theta=0^{\circ}$ and $\phi=90^{\circ}$ mean $+y$-direction, and $\theta=90^{\circ}$ means $+z$-direction. Note that the impact site in this figure is $\phi = 180^{\circ}$ and $\theta = 0^{\circ}$.}
    \label{20kmSurfaceProfile-largestBody-model1-30-3.25kms}
  \end{center}
\end{figure} 
Figure \ref{20kmSurfaceProfile-largestBody-model1-30-3.25kms} shows the mass fractions of materials (metals, mantle, and crust) on the surface of the resultant body shown in Fig.\,\ref{subsequent-pictures-model1-30-3.25kms}f. Here, ``surface'' materials mean materials that are located at a depth of less than $20\,{\rm km}$. Note that our simulations resolve $20\,{\rm km}$ as 2--3 SPH particles, and thus SPH particles that are located at a depth of less than $20\,{\rm km}$ represent surface materials. Fig.\,\ref{20kmSurfaceProfile-largestBody-model1-30-3.25kms}a shows that the crust fraction is small around the longitude of $\phi=180^{\circ}$ and latitude of $\theta=0^{\circ}$, which corresponds to the impact site. Mantle materials are mainly exposed at the impact site, and the average value of the mantle fraction over the entire surface is about 50\% (Fig.\,\ref{20kmSurfaceProfile-largestBody-model1-30-3.25kms}b). Metallic materials are broadly distributed around the impact site and surface metals exist not only at sites with a small crust fraction, but also at sites with a large crust fraction (Fig.\,\ref{20kmSurfaceProfile-largestBody-model1-30-3.25kms}c). For example, the surface materials at $\phi = 70^{\circ}$ and $\theta = 60^{\circ}$ are composed of $\approx 90$\% crust and a small amount of metals. The metallic materials at sites with a large crust fraction correspond to core materials that were ejected and reaccumulated on the original surface crust (Fig.\,\ref{subsequent-pictures-model1-30-3.25kms}d, e). Mesosiderites may form at such sites with a large crust fraction and small amounts of metals. Note that mesosiderites are composed of similar amounts of silicates and metals, so that it seems that sites with large crust fractions but small metal fractions are not appropriate sources for mesosiderite-like materials; that is, the crust fraction is too much larger than the metal fraction. However, the fraction is the average value across the resolutions of our simulations, or $\sim10\,{\rm km}$. If we observe produced materials at meteorite scales of $\sim1$--$100\,{\rm cm}$ and metals are heterogeneously distributed, there may be mesosiderite-like materials with similar amounts of silicates and metals.

\begin{figure}[!htb]
  \begin{center}
    \includegraphics[bb=0 0 865 540, width=1.0\linewidth,clip]{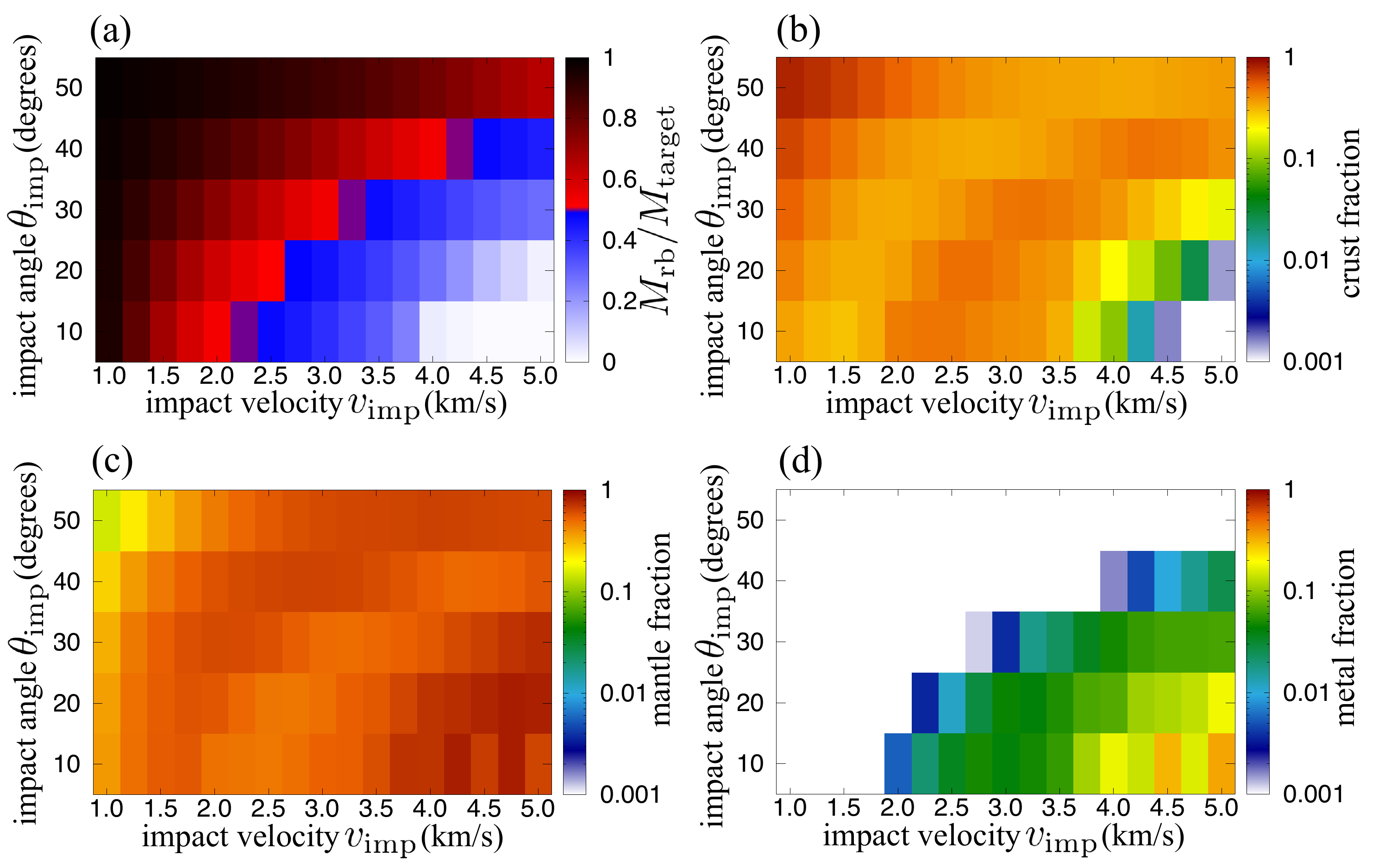}
    \caption{Properties of the resultant bodies produced through various giant impact simulations with model 1. The horizontal axes show the impact velocities $v_{{\rm imp}}$, and the vertical axes show the impact angles $\theta_{{\rm imp}}$. Panel (a) shows the mass of the resultant body $M_{{\rm rb}}$ produced by each combination of impact velocity and angle. Panels (b), (c), and (d) show the fractions of crust, mantle, and metals, respectively, across the entire surface of the resultant body produced by each impact combination. As before, the surface means materials at a depth of less than $20\,{\rm km}$.}
    \label{parameter-survey-collision-to-Vesta-model1}
  \end{center}
\end{figure}

Figure \ref{parameter-survey-collision-to-Vesta-model1} shows the mass and surface properties of the resultant bodies produced by impact simulations with model 1 with various combinations of impact velocity $v_{{\rm imp}}$ and angle $\theta_{{\rm imp}}$. Hereafter, $M_{{\rm rb}}$ shows the mass of the resultant body. The impacts with low impact velocities and high impact angles that result in $M_{{\rm rb}}/M_{{\rm target}}>0.5$ only eject crust and mantle materials, and thus do not carry metallic materials to the surfaces of the resultant bodies: catastrophic collisions with $M_{{\rm rb}}/M_{{\rm target}}<0.5$ are necessary to transport the metal core materials to the surfaces. The fraction of surface mantle is $\sim50$\% even for the impacts with $M_{{\rm rb}}/M_{{\rm target}} \approx 0.5$, and more destructive collisions result in a larger fraction of surface mantle and metals, which are difficult to retain the original crust and thus are very unpromising for the mesosiderite formation on the surfaces.

\begin{figure}[!htb]
  \begin{center}
    \includegraphics[bb=0 0 461 256, width=1.0\linewidth,clip]{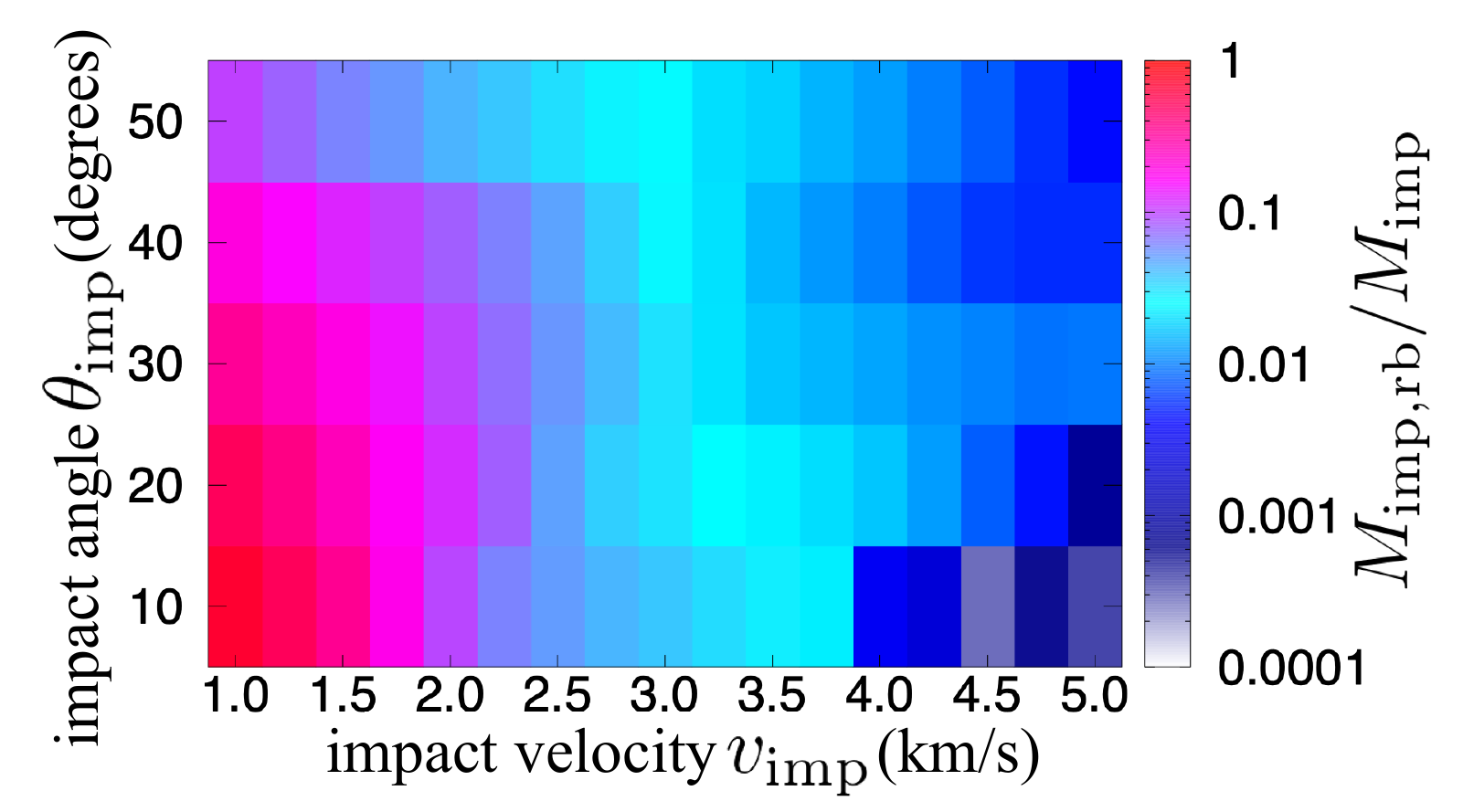}
    \caption{Fractions of impactors that are incorporated into resultant bodies through various giant impact simulations with model 1. The horizontal axis shows the impact velocity $v_{{\rm imp}}$, and the vertical axis shows the impact angle $\theta_{{\rm imp}}$. The color shows the mass of an impactor that is incorporated into a resultant body $M_{{\rm imp,rb}}$ normalized by the mass of the original impactor $M_{{\rm imp}}$.}
    \label{impactor-fraction-inforporated-into-target}
  \end{center}
\end{figure}

As we discussed in Section \ref{Introduction}, we argue that the mixing of a metal core and a crust of the same parent body through a giant impact is a likely mechanism for the mesosiderite formation and we focus on it. However, mixing of a metal core of an impactor and a basaltic crust of a target has long been considered as a way to make mesosiderites (e.g.,\,\citealt{Hassanzadeh-et-al1990}). Moreover, coalescence of a basaltic impactor and a target body can thicken a basaltic crust of a target. Therefore, the investigation of how an impactor is incorporated into a target body is a valuable work and our simulations can provide some information for this topic. Fig.\,\ref{impactor-fraction-inforporated-into-target} shows how much the impactors are incorporated into the target body for model 1 impacts. Considerable amounts of more than 10\% of the original impactor are incorporated into the target body for the impacts with $v_{{\rm imp}} < 2\,{\rm km/s}$ and $\theta_{{\rm imp}} < 40^{\circ}$, and it may be possible that the mixing of a metal core of an impactor and a crust of a target for such impacts. However, based on the analysis of the impact velocities expected in the present main asteroid belt, the frequency of impacts with impact velocities $< 2\,{\rm km/s}$ is about 9\%, while that with $2\,{\rm km/s} <$ impact velocities $< 5\,{\rm km/s}$ is about 50\% (\citealt{Sugiura2020}). Thus, an impact like Fig.\,\ref{subsequent-pictures-model1-30-3.25kms} more frequently occurs than coalescence of a target and an impactor.

\subsection{Model 2: 140\,km core radius and 80\,km crust thickness\label{model2}}

\begin{figure}[!htb]
  \begin{center}
    \includegraphics[bb=0 0 1063 540, width=1.0\linewidth,clip]{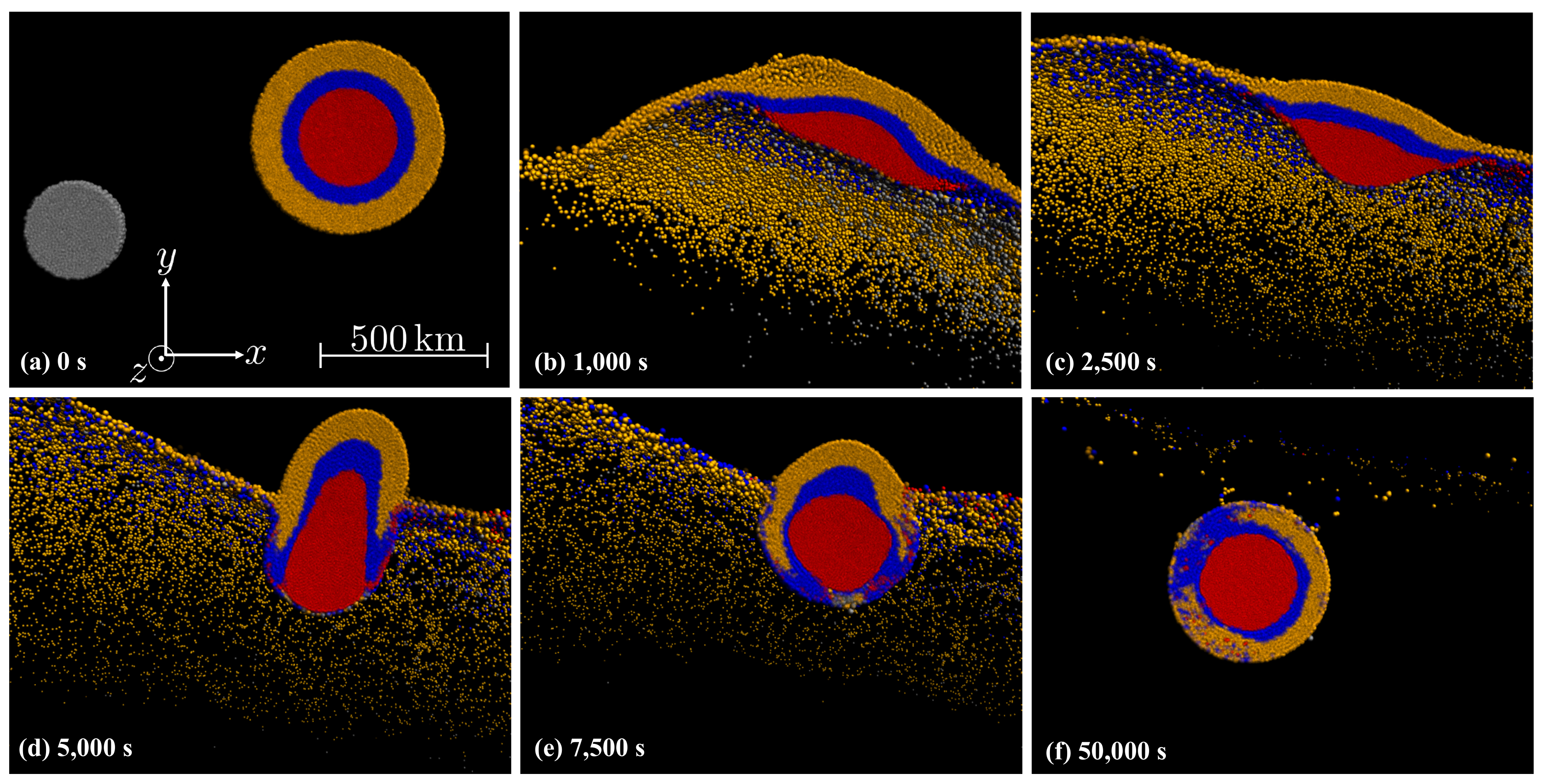}
    \caption{Snapshots of a simulation with model 2: giant impact with $\theta_{{\rm imp}}=40^{\circ}$ and $v_{{\rm imp}}=4.0\,{\rm km/s}$ onto a differentiated body with a core radius of 140 km and crust thickness of 80 km. Similar to Fig.\,\ref{subsequent-pictures-model1-30-3.25kms}, grey SPH particles are almost invisible because they are lost to space.}
    \label{subsequent-pictures-model2-40-4.0kms}
  \end{center}
\end{figure}

Figure \ref{subsequent-pictures-model2-40-4.0kms} shows snapshots of an impact simulation with model 2, that is, a simulation with a differentiated target body having a 140\,km core radius and 80\,km crust thickness. We set $\theta_{{\rm imp}}=40^{\circ}$ and $v_{{\rm imp}}=4.0\,{\rm km/s}$ for the simulation. Similar to Fig.\,\ref{subsequent-pictures-model1-30-3.25kms}, the metal core was excavated by the impact and metal materials were ejected (Fig.\,\ref{subsequent-pictures-model2-40-4.0kms}b, c). However, the mass of the resultant body was $M_{{\rm rb}} \approx 0.7 M_{{\rm target}}$, demonstrating that an even less destructive impact compared to Fig.\,\ref{subsequent-pictures-model1-30-3.25kms} can excavate the metal core. The mantle layer was exposed only at the impact site, and a wide area over the surface retained the original crust layer (Fig.\,\ref{subsequent-pictures-model2-40-4.0kms}f). Thus, most of the ejected metals accreted on the crust layer.

\begin{figure}[!htb]
  \begin{center}
    \includegraphics[bb=0 0 886 546, width=1.0\linewidth,clip]{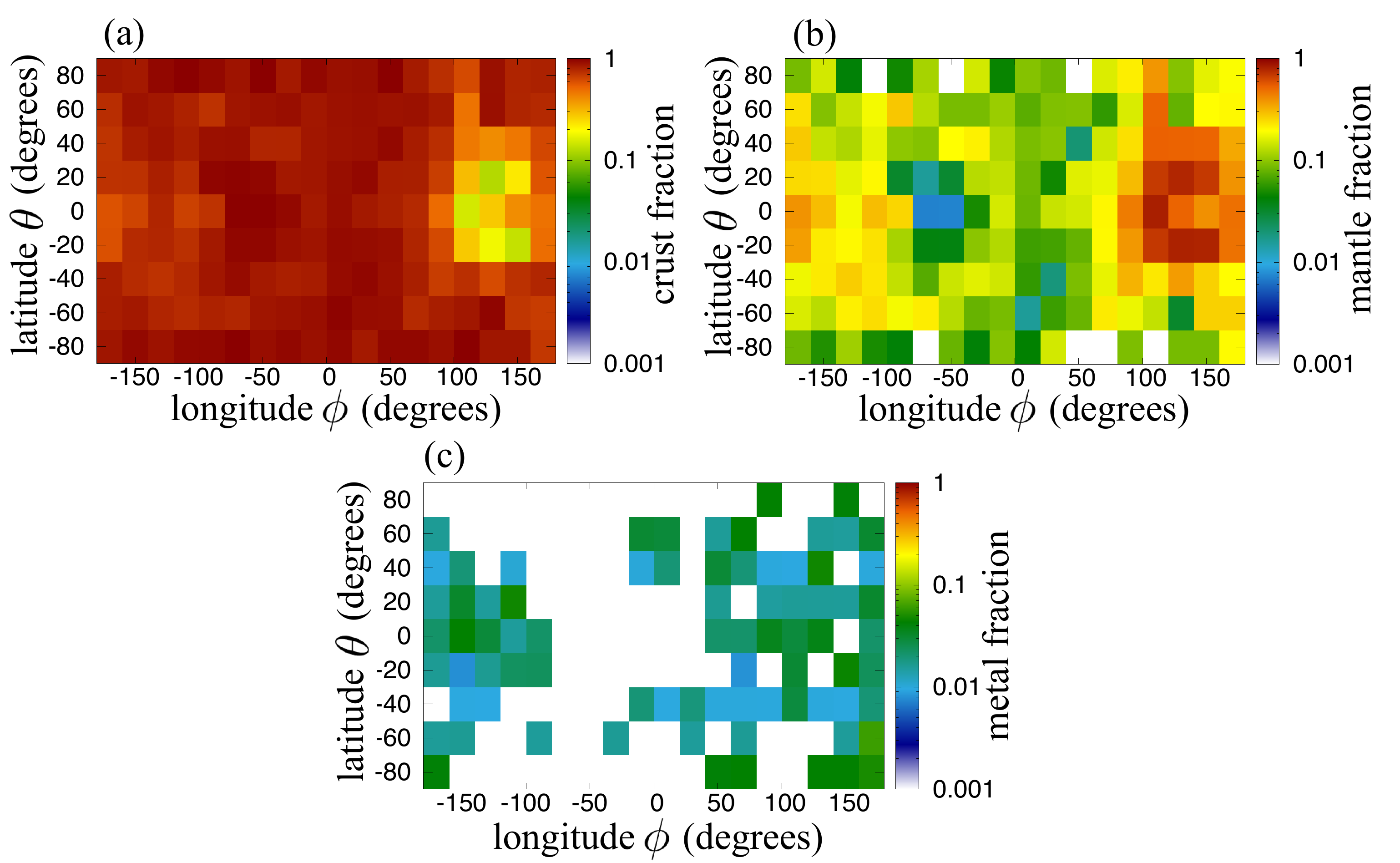}
    \caption{Mass fractions of materials on the surface of the resultant body for model 2: (a) crust, (b) mantle, and (c) metals. This surface was produced through a giant impact with $\theta_{{\rm imp}}=40^{\circ}$ and $v_{{\rm imp}}=4.0\,{\rm km/s}$ onto a differentiated body with a core radius of 140\,km and crust thickness of 80\,km. The fraction of the surface materials is measured at $t=5.0\times 10^{4}\,{\rm s}$ (Fig.\,\ref{subsequent-pictures-model2-40-4.0kms}f). Note that the impact site in this figure is $\phi = 140^{\circ}$ and $\theta = 0^{\circ}$.}
    \label{20kmSurfaceProfile-largestBody-model2-40-4.0kms}
  \end{center}
\end{figure}

Figure \ref{20kmSurfaceProfile-largestBody-model2-40-4.0kms} shows the mass fractions of materials (metals, mantle, and crust) on the surface of the resultant body shown in Fig.\,\ref{subsequent-pictures-model2-40-4.0kms}f. Except for the impact site at $\phi = 140^{\circ}$ and $\theta = 0^{\circ}$, a large area of the surface is covered with materials with a crust fraction of $>90$\% (Fig.\,\ref{20kmSurfaceProfile-largestBody-model2-40-4.0kms}a). Even the average value of the crust fraction for the entire surface of the resultant body is $\approx 75$\%. Moreover, small amounts of metal materials are spread over a broad area of the surface (Fig.\,\ref{20kmSurfaceProfile-largestBody-model2-40-4.0kms}c). Thus, except for the impact site, the formed materials are mainly composed of crust and also contain small amounts of metals. Mesosiderite-like materials may form at microscopic points within such materials.

\begin{figure}[!htb]
  \begin{center}
    \includegraphics[bb=0 0 865 540, width=1.0\linewidth,clip]{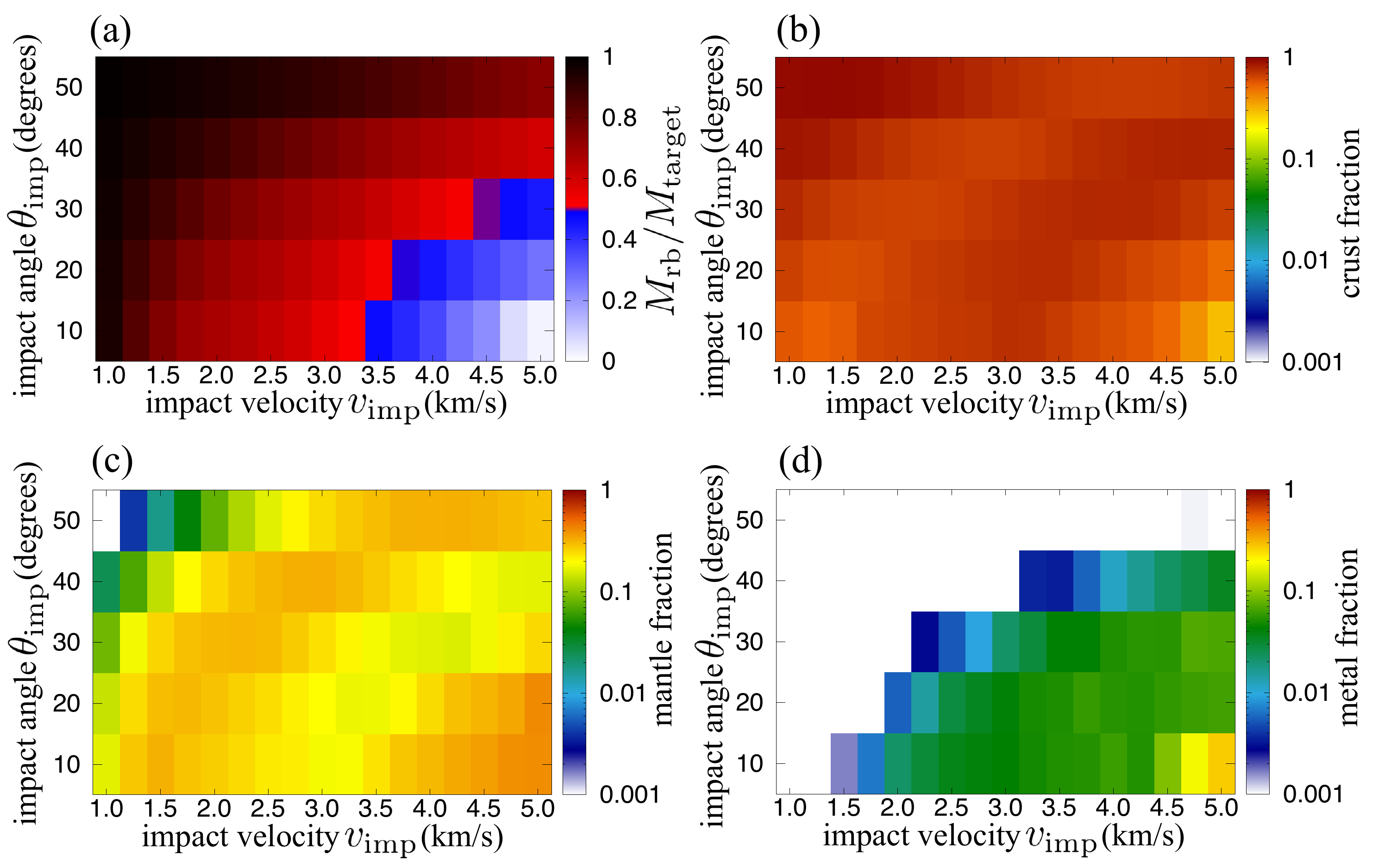}
    \caption{Properties of the resultant bodies produced through the various giant impact simulations with model 2. The horizontal axes show the impact velocities $v_{{\rm imp}}$, and the vertical axes show the impact angles $\theta_{{\rm imp}}$. Panel (a) shows the mass of the resultant body $M_{{\rm rb}}$ produced by each combination of impact velocity and angle. Panels (b), (c), and (d) show the fractions of crust, mantle, and metals, respectively, over the entire surface of the resultant body produced by each impact combination.}
    \label{parameter-survey-collision-to-Vesta-model2}
  \end{center}
\end{figure}

Figure \ref{parameter-survey-collision-to-Vesta-model2} is similar to Fig.\,\ref{parameter-survey-collision-to-Vesta-model1}, but shows the results of parameter surveys for model 2. The larger core compared to model 1 allows relatively less destructive impacts with $M_{{\rm rb}} / M_{{\rm target}} \approx 0.7$ to transport metal core materials to the surface. For such impacts, $\sim 70\%$ of the entire surface of the resultant bodies is covered with crust materials, so that mantle materials are not as exposed on the surfaces as we see in Fig.\,\ref{subsequent-pictures-model2-40-4.0kms}. Thus, even the impacts that excavate the metal core leave a surface that mostly retains the original crust. Although favored impact velocities for such impacts depend on impact angles, we roughly require $1.5\,{\rm km/s} \lesssim v_{{\rm imp}} \lesssim 5.0\,{\rm km/s}$ to excavate the metal core with retaining the original crust. Such impacts with this range of impact velocities frequently occur in the main belt (\citealt{Sugiura2020}).

\section{Discussion \label{Discussion}}
For model 1 (core radius of $110\,{\rm km}$ and crust thickness of $40\,{\rm km}$), the impacts that resulted in $M_{{\rm rb}} \approx 0.5 M_{{\rm target}}$ excavated the metal core, and mantle materials were exposed over almost half of the surface area. Although mantle materials were exposed over large areas on the surface, it is possible that some surface locations locally formed with a large crust fraction, $>90\%$, and a small amount of metal materials. For model 2 (core radius of $140\,{\rm km}$ and crust thickness of $80\,{\rm km}$), the relatively less destructive impacts with $M_{{\rm rb}} \approx 0.7 M_{{\rm target}}$ still excavated the metal core but exposed mantle materials only at the impact sites. Large areas of the surface retained the original crust layers, and thus the re-accumulation of ejected metal materials on the surfaces mainly created materials with a large crust fraction and small metal fraction. Especially for model 2, relatively non-destructive impacts excavated the metal core without not largely changing the initial internal structure.

For model 1, mantle materials were exposed over large areas on the surface and the mantle fraction for the entire surface was $\sim50\%$. Thus stony-iron materials produced at the surface are likely to contain mantle materials with large fractions, $>10\%$, which is inconsistent with the composition of mesosiderites. Moreover, stony-iron materials that are mainly composed of mantle and metal materials probably formed on the surfaces. We already found pallasite meteorites as an example of such materials with a mixture of mantle and metals, but the oxygen isotope ratio of pallasites is distinctly different from that of mesosiderites; that is, the parent body of pallasites is probably different from that of mesosiderites (\citealt{Greenwood-et-al2006}). Thus, pallasite-like materials should not be produced on the surface of the parent body of mesosiderites, and model 1 is not conducive to the formation of mesosiderite-like materials without the formation of pallasite-like materials.

Pallasites are usually considered to form near the core-mantle boundary through intrusion of molten metal from a core into a surrounding olivine mantle (e.g.,\,\citealt{Scott1977}). Our simulations show that even impacts that retain the internal structure of asteroid transport materials from the core-mantle boundary to the surface. Because surface materials can be ejected by cratering impacts, a giant impact is one possible mechanism that contributes to bringing materials from a core-mantle boundary to the Earth. In contrast, it is suggested that some pallasites experienced rapid cooling after their formation (e.g.,\,\citealt{Miyamoto-and-Takeda1993, Hsu2003}), and such rapid cooling is difficult to realize near a core-mantle boundary. The mixing of metal and mantle materials on surfaces can realize such rapid cooling, so that the giant impacts shown in our simulations may be one possible mechanism for the formation of pallasites.

\begin{table}[!htb]
  \begin{center}
    \begin{tabular}{c c c c c}\hline 
      $\phi (^\circ)$ & $\theta (^\circ)$ & metal fraction (\%) & mantle fraction (\%) & crust fraction (\%) \rule[0mm]{0mm}{5mm}\\
      \hline
      70 & 60 & 4.4 & 5.8 & 89.8 \rule[0mm]{0mm}{5mm}\\
      50 & 40 & 3.0 & 2.0 & 95.0 \rule[0mm]{0mm}{5mm}\\
      10 & -40 & 1.0 & 5.8 & 93.2 \rule[0mm]{0mm}{5mm}\\
      30 & -40 & 1.9 & 1.9 & 96.2 \rule[0mm]{0mm}{5mm}\\
      50 & -80 & 4.2 & 0.0 & 95.8 \rule[0mm]{0mm}{5mm}\\
      70 & -80 & 4.4 & 0.0 & 95.6 \rule[0mm]{0mm}{5mm}\\
      \hline
    \end{tabular}
  \end{center}
  \caption{Fractions of the three types of materials at the most promising positions for the mesosiderite formation in Fig.\,\ref{20kmSurfaceProfile-largestBody-model2-40-4.0kms}, i.e., positions with mantle fractions $<6\%$ and small amounts of metals.}
  \label{fractions-at-promising-positions-for-mesosiderite-formation}
\end{table}

For model 2, large surface areas, except at the impact site, retained the original crust with a crust fraction of $>90\%$, which reduced the possibility of the formation of pallasite-like materials. Some surface positions, such as $\phi = 50^{\circ}$ and $\theta = 40^{\circ}$ in Fig.\,\ref{20kmSurfaceProfile-largestBody-model2-40-4.0kms}, showed materials with a mantle fraction of $<6\%$ and small amounts of metals. These positions are considered to be the most promising sites for the mesosiderite formation. Fractions of materials at these positions are tabulated in Table \ref{fractions-at-promising-positions-for-mesosiderite-formation}. At these positions, mantle fractions are $<6\%$ and thus produced materials are likely to have such small amounts of mantle materials. Although the crust fractions $\gg$ the metal fractions, the fractions shown are average values across numerical pixel scales. If metal materials are heterogeneously distributed within these pixels, some microscopic points with meteorite scales may have crust fraction $\sim$ metal fraction, which can explain the composition of mesosiderites. Although the surface area with such a small fraction of mantle is small, $\sim1\%$ of the entire surface, the formation of mesosiderite-like materials on the surface is possible. Moreover, the in-situ observation of the asteroid Vesta by the Dawn probe did not clearly detect mesosiderite-like materials on its surface (the fractions of mesosiderite-like materials $<10\%$), so no macroscopic locations for the formation sites of mesosiderites are apparent on Vesta's surface (\citealt{Peplowski-et-al2013}). This is consistent with our simulation results, where the mesosiderite-like material formation was very sparse and the average fractions of mesosiderite-like materials $<10\%$ (Fig.\,\ref{20kmSurfaceProfile-largestBody-model2-40-4.0kms} and Table \ref{fractions-at-promising-positions-for-mesosiderite-formation}). Therefore, model 2, with an internal structure having a 140\,km core radius and 80\,km crust thickness, is more conducive to the formation of mesosiderites.

\begin{figure}[!htb]
  \begin{center}
    \includegraphics[bb=0 0 1063 540, width=1.0\linewidth,clip]{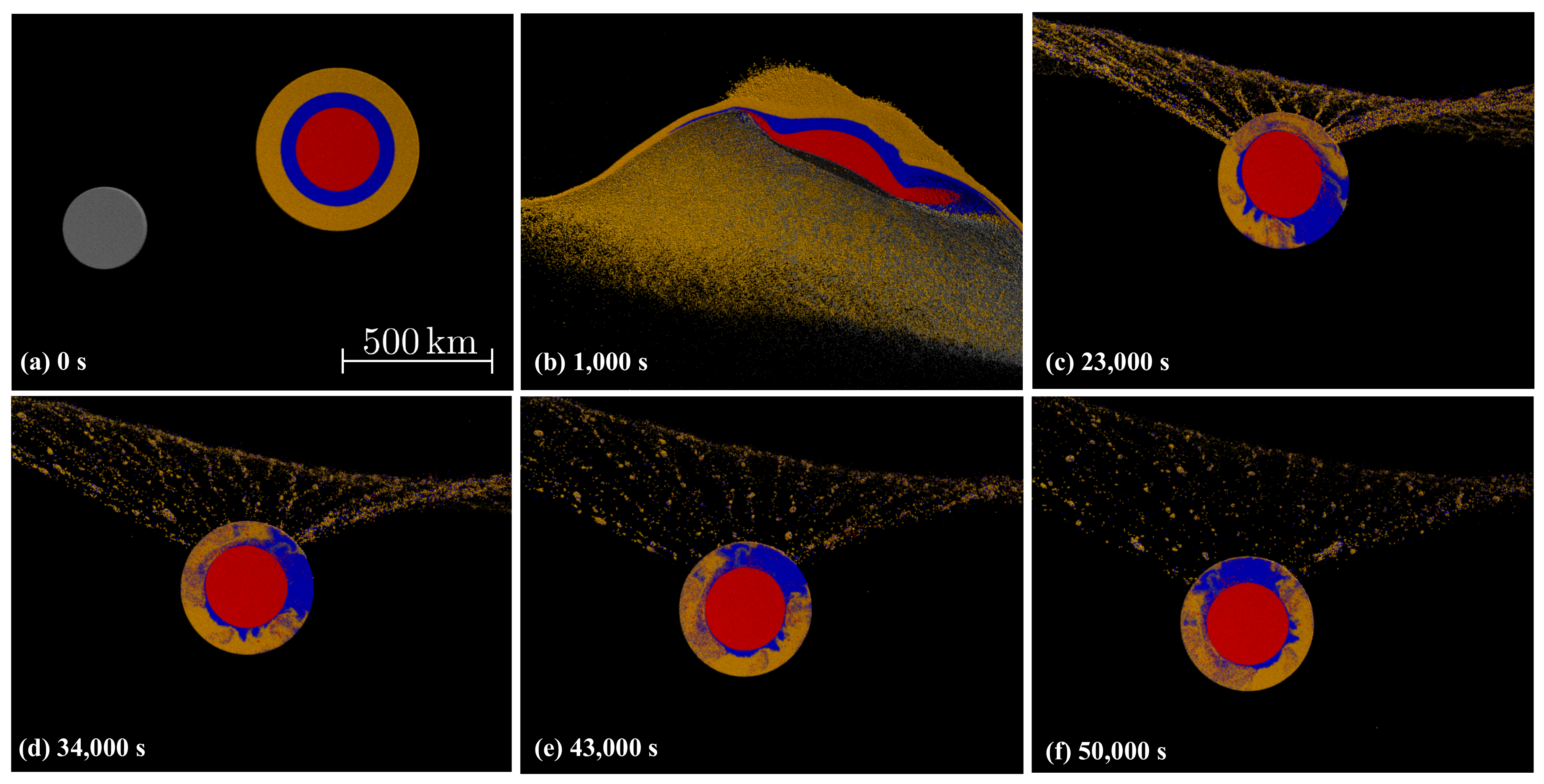}
    \caption{Snapshots of a high-resolution simulation with $6\times 10^{6}$ SPH particles of a giant impact with $\theta_{{\rm imp}}=40^{\circ}$ and $v_{{\rm imp}}=4.0\,{\rm km/s}$, onto a differentiated body with a core radius of 140 km and crust thickness of 80 km (model 2).}
    \label{subsequent-pictures-model2-40-4.0kms-Ntotal=6e6}
  \end{center}
\end{figure}

If we consider that the asteroid Vesta is a parent body of mesosiderites, we need to discuss the comparison between the resultant body produced through our simulation and the actual asteroid. The Dawn probe investigated the detailed surface state of Vesta and found very few mantle-rich sites on its surface (\citealt{Nathues-et-al2015}). In contrast, our simulation shows that, even for model 2, mantle materials are exposed at the impact site (see Fig.\,\ref{subsequent-pictures-model2-40-4.0kms}f), which is not consistent with the observed surface state of Vesta. However, we additionally conducted a higher-resolution simulation with model 2, as shown in Fig.\,\ref{subsequent-pictures-model2-40-4.0kms}, and found that exposed mantle possibly can be covered by ejected crust materials. Fig.\,\ref{subsequent-pictures-model2-40-4.0kms-Ntotal=6e6} shows snapshots of the simulation with the same impact condition as Fig.\,\ref{subsequent-pictures-model2-40-4.0kms} but with a much larger number of SPH particles, $\approx 6\times 10^{6}$. The impact mainly ejected surface crust (Fig.\,\ref{subsequent-pictures-model2-40-4.0kms-Ntotal=6e6}b). Mantle materials were initially exposed at the impact site (Fig.\,\ref{subsequent-pictures-model2-40-4.0kms-Ntotal=6e6}c), but subsequent accretion of the ejected crust materials covered the exposed mantle (Fig.\,\ref{subsequent-pictures-model2-40-4.0kms-Ntotal=6e6}d-f). This result is consistent with the present surface state of Vesta.  

\section{Summary \label{Summary}}
Mesosiderites are a type of stony-iron meteorite composed of silicates and Fe-Ni metals. The mesosiderite silicates and metals respectively originated from the crust and metal core of a differentiated asteroid. However, mesosiderites rarely contain olivine, which is mainly found in mantle layers. One possible scenario for the formation of mesosiderites is that a giant impact onto a differentiated asteroid excavated its metal core and transported metal material to the surface crust, where it mixed with crust materials. However, it is not obvious that such a giant impact, which does not cause catastrophic disruption of a differentiated asteroid or its internal structure, can produce materials that contain crust and metal materials but not mantle materials.

We conducted three-dimensional numerical simulations using the SPH method and investigated how a giant impact onto a differentiated asteroid excavates and mixes internal materials. We considered two types of internal structures of differentiated target bodies. One of them is based on the magma-ocean crystallization model of the asteroid Vesta (\citealt{Mandler-and-ElkinsTanton2013}), which is a likely parent body of mesosiderites, and we set the radius of the target body, the radius of the metal core, the thickness of the mantle layer, and the thickness of the crust layer as $270\,{\rm km}$, $110\,{\rm km}$, $120\,{\rm km}$, and $40\,{\rm km}$, respectively (model 1). As another possible internal structure, we conducted giant impact simulations with a target body with a metal core radius of $140\,{\rm km}$, a mantle thickness of $50\,{\rm km}$, and a crust thickness of $80\,{\rm km}$ (model 2), which considered the suggestions obtained from the Dawn observation of Vesta (\citealt{Clenet-et-al2014, Ermakov-et-al2014}). We used both models to conduct impact simulations with various impact velocities and impact angles.

As a result, we found that, for model 1, excavation of the metal core needs a nearly catastrophic impact that results in the masses of the resultant bodies $M_{{\rm rb}}$ becoming 0.5 times that of the original target body $M_{{\rm target}}$ (Fig.\,\ref{parameter-survey-collision-to-Vesta-model1}). For such impacts, although a very limited number of surface locations of the resultant bodies have materials with a large fraction of crust materials, $>90\%$, and small amounts of metal materials (Fig.\,\ref{20kmSurfaceProfile-largestBody-model1-30-3.25kms}), mantle materials are exposed over almost half the area of the entire surface (Fig.\,\ref{subsequent-pictures-model1-30-3.25kms}). Thus, stony-iron materials produced in this way are likely to contain significant amounts of mantle materials, and it is difficult to produce mesosiderite-like materials. Moreover, since mantle materials are largely exposed on the surface, produced stony-iron materials may be composed of only mantle and metal materials, i.e., pallasite-like materials. Thus the thin-crust model may be preferable for the formation of pallasites on the surface of a pallasite parent body through a giant impact.

In contrast, for model 2, relatively non-destructive impacts with $M_{{\rm rb}} \approx 0.7 M_{{\rm target}}$ excavate the metal core (Fig.\,\ref{parameter-survey-collision-to-Vesta-model2}). For such impacts, mantle materials are exposed only at impact site, and other large surface areas retain the original crust (Fig.\,\ref{subsequent-pictures-model2-40-4.0kms}). Thus, excavated metal materials mainly accrete on crust-dominated surfaces. Several positions on the surface have crust-dominated materials with a mantle fraction of $<6\%$ and small amounts of metal materials (Fig.\,\ref{20kmSurfaceProfile-largestBody-model2-40-4.0kms}). Although even at these promising positions crust fraction $\gg$ mantle fraction $\sim$ metal fraction, which seems to be inconsistent with mesosiderite compositions, the fraction shown in our study is average fraction across numerical resolution $\sim 10\,{\rm km}$ and fractions at some microscopic points with meteorite scales $\sim 1$--$100\,{\rm cm}$ may be crust fraction $\sim$ metal fraction $\gg$ mantle fraction if materials are heterogeneously distributed within such positions. Thus, mesosiderite formation is possible for model 2.

We conclude that the thin-crust model (model 1) inferred from conventional magma-ocean crystallization model makes it difficult to explain the formation of mesosiderites, but in contrast the thick-crust model (model 2) is conducive to the formation of mesosiderites. It is suggested that post-magma ocean volcanism may make a crust thicker than $60\,{\rm km}$ (\citealt{Yamaguchi-et-al2011}), which can explain the thick crust of our model 2. Our simulation results may constrain the internal structure of a mesosiderite parent body and its differentiation process.

\section*{Acknowledgements}
K.S. and H.G. acknowledge the financial support of the Ministry of Education, Culture, Sports, Science and Technology, KAKENHI Grant (JP17H06457). H.G. acknowledges the financial support of the Japan Society for the Promotion of Science (JSPS), KAKENHI Grant (JP17H02990). M.K.H. acknowledges the financial support of JSPS KAKENHI Grant (JP19K03946). K.S. acknowledges the financial support of JSPS KAKENHI Grant (JP20K14536, JP20J01165). Numerical simulations in this work were carried out on the Cray XC50 supercomputer at the Center for Computational Astrophysics, National Astronomical Observatory of Japan.

\appendix
\section{Equations used in simulations}
In our impact simulations, we solve the following equations of continuity, motion, and energy:

\begin{align}
  & \frac{d\rho_{i}}{dt} = \sum_{j}m_{j}(\bm{v}_{i}-\bm{v}_{j})\cdot \frac{\partial}{\partial \bm{x}_{i}}W(|\bm{x}_{i}-\bm{x}_{j}|,h_{i}), \label{EoC} \\
  & \frac{d\bm{v}_{i}}{dt} = -\sum_{j} m_{j} \Bigl[ \frac{p_{i}}{\rho_{i}^{2}} + \frac{p_{j}}{\rho_{j}^{2}} + \Pi_{ij} \Bigr] \frac{\partial}{\partial \bm{x}_{i}}W(|\bm{x}_{i}-\bm{x}_{j}|,0.5[h_{i}+h_{j}]) + \sum_{j}\bm{g}_{{\rm grav},ij}, \label{EoM} \\
  & \frac{du_{i}}{dt} = \frac{1}{2} \sum_{j}m_{j} \Bigl[ \frac{p_{i}}{\rho_{i}^{2}} + \frac{p_{j}}{\rho_{j}^{2}} + \Pi_{ij} \Bigr] (\bm{v}_{i}-\bm{v}_{j})\cdot \frac{\partial}{\partial \bm{x}_{i}}W(|\bm{x}_{i}-\bm{x}_{j}|,0.5[h_{i}+h_{j}]), \label{EoE}
\end{align}

\noindent where the subscripts of Roman letters are the indices of SPH particles; $\rho$, $p$, and $u$ are density, pressure, and specific internal energy at the positions of SPH particles, respectively; $\bm{x}$, $\bm{v}$, and $m$ are the positions, velocities, and masses of SPH particles, respectively; $W(r, h)$ is the cubic spline kernel function and $h$ represents its smoothing length; $\bm{g}_{{\rm grav},ij}$ is the mutual gravity between the $i$-th and $j$-th SPH particles; and $\Pi_{ij}$ is an artificial viscosity term. For $\Pi_{ij}$, we used the standard Monaghan artificial viscosity with $\alpha=1$ and $\beta=2$ (e.g.,\,\citealt{Monaghan1992}). We did not apply the Balsara switch and any types of time dependent viscosity terms. Note that we used $h_{i}$ for the kernel function in Eq.\,(\ref{EoC}) because so-called ``gather'' formulation of the density representation gives more appropriate density distribution around contact discontinuities (see \citealt{Inutsuka2002}), while we used $0.5[h_{i}+h_{j}]$ for Eqs. (\ref{EoM}) and (\ref{EoE}) because of the conservation of the linear momentum and energy.

We solve the equation of continuity, or the time evolution equation of the density Eq.\,(\ref{EoC}). The initial densities of SPH particles that compose metallic cores and rocky layers are set to uncompressed densities of iron and basalt calculated from their Tillotson equations of state, respectively. In our simulations, all SPH particles have the same mass. Thus, for consistency between the initial densities for metallic cores and rocky layers, the initial particle spacing for SPH particles in metallic cores is smaller than that for SPH particles in rocky layers. The smoothing length of each SPH particle is fixed to the initial particle spacing at its neighbor, and thus the smoothing length of a core particle is smaller than that of a crust or a mantle particle. FDPS guarantees the symmetry of the interactions between all SPH particles, so that the linear momentum is completely conserved within the rounding error even if we use different values of smoothing lengths for different SPH particles. We also confirmed that the use of variable smoothing length does not change results. The initial specific internal energy of each SPH particle is set to $5.0\times 10^{9}\,{\rm erg/g}$.

\bibliography{mybibfile}

\begin{thebibliography}{}

\bibitem[\protect\citename{{Asphaug} {\em et~al.}, }2006]{Asphaug-et-al2006}
{Asphaug}, E., {Agnor}, C.~B., \& {Williams}, Q. 2006.
\newblock {Hit-and-run planetary collisions}.
\newblock {\em \nat}, {\bf 439}(7073), 155--160.

\bibitem[\protect\citename{{Benz} \& {Asphaug}, }1999]{Benz-and-Asphaug1999}
{Benz}, W., \& {Asphaug}, E. 1999.
\newblock {Catastrophic Disruptions Revisited}.
\newblock {\em \icarus}, {\bf 142}(Nov.), 5--20.

\bibitem[\protect\citename{{Bercovici} {\em et~al.},
  }2019]{Bercovici-et-al2019}
{Bercovici}, H.~L., {Elkins-Tanton}, L.~T., \& {Schaefer}, L. 2019 (Mar.).
\newblock {The Effect of Bulk Composition on the Behavior of Sulfur During Core
  Formation}.
\newblock {\em Page  1366 of:} {\em 50th Annual Lunar and Planetary Science
  Conference}.
\newblock Lunar and Planetary Science Conference.

\bibitem[\protect\citename{{Carter} {\em et~al.}, }2018]{Carter-et-al2018}
{Carter}, P.~J., {Leinhardt}, Z.~M., {Elliott}, T., {Stewart}, S.~T., \&
  {Walter}, M.~J. 2018.
\newblock {Collisional stripping of planetary crusts}.
\newblock {\em \epsl}, {\bf 484}(Feb), 276--286.

\bibitem[\protect\citename{{Clenet} {\em et~al.}, }2014]{Clenet-et-al2014}
{Clenet}, H., {Jutzi}, M., {Barrat}, J.-A., {Asphaug}, E.~I., {Benz}, W., \&
  {Gillet}, P. 2014.
\newblock {A deep crust-mantle boundary in the asteroid 4 Vesta}.
\newblock {\em \nat}, {\bf 511}(7509), 303--306.

\bibitem[\protect\citename{{Emsenhuber} {\em et~al.},
  }2018]{Emsenhuber-et-al2018}
{Emsenhuber}, A., {Jutzi}, M., \& {Benz}, W. 2018.
\newblock {SPH calculations of Mars-scale collisions: The role of the equation
  of state, material rheologies, and numerical effects}.
\newblock {\em \icarus}, {\bf 301}(Feb.), 247--257.

\bibitem[\protect\citename{{Ermakov} {\em et~al.}, }2014]{Ermakov-et-al2014}
{Ermakov}, A.~I., {Zuber}, M.~T., {Smith}, D.~E., {Raymond}, C.~A., {Balmino},
  G., {Fu}, R.~R., \& {Ivanov}, B.~A. 2014.
\newblock {Constraints on Vesta{\textquoteright}s interior structure using
  gravity and shape models from the Dawn mission}.
\newblock {\em \icarus}, {\bf 240}(Sep), 146--160.

\bibitem[\protect\citename{{Farinella} \& {Davis},
  }1992]{Farinella-and-Davis1992}
{Farinella}, Paolo, \& {Davis}, Donald~R. 1992.
\newblock {Collision rates and impact velocities in the main asteroid belt}.
\newblock {\em \icarus}, {\bf 97}(1), 111--123.

\bibitem[\protect\citename{{Genda} {\em et~al.}, }2012]{Genda-et-al2012}
{Genda}, H., {Kokubo}, E., \& {Ida}, S. 2012.
\newblock {Merging Criteria for Giant Impacts of Protoplanets}.
\newblock {\em \apj}, {\bf 744}(Jan.), 137.

\bibitem[\protect\citename{{Gingold} \& {Monaghan},
  }1977]{Gingold-and-Monaghan1977}
{Gingold}, R.~A., \& {Monaghan}, J.~J. 1977.
\newblock {Smoothed particle hydrodynamics - Theory and application to
  non-spherical stars}.
\newblock {\em \mnras}, {\bf 181}(Nov.), 375--389.

\bibitem[\protect\citename{{Golabek} {\em et~al.}, }2018]{Goblack-et-al2018}
{Golabek}, G.~J., {Emsenhuber}, A., {Jutzi}, M., {Asphaug}, E.~I., \& {Gerya},
  T.~V. 2018.
\newblock {Coupling SPH and thermochemical models of planets: Methodology and
  example of a Mars-sized body}.
\newblock {\em \icarus}, {\bf 301}(Feb.), 235--246.

\bibitem[\protect\citename{{Greenwood} {\em et~al.},
  }2006]{Greenwood-et-al2006}
{Greenwood}, R.~C., {Franchi}, I.~A., {Jambon}, A., {Barrat}, J.~A., \&
  {Burbine}, T.~H. 2006.
\newblock {Oxygen Isotope Variation in Stony-Iron Meteorites}.
\newblock {\em \science}, {\bf 313}(5794), 1763--1765.

\bibitem[\protect\citename{{Haack} {\em et~al.}, }1990]{Haack-et-al1990}
{Haack}, H., {Rasmussen}, K.~L., \& {Warren}, P.~H. 1990.
\newblock {Effects of regolith/megaregolith insulation on the cooling histories
  of differentiated asteroids.}
\newblock {\em \jgr}, {\bf 95}(Apr.), 5111--5124.

\bibitem[\protect\citename{{Haack} {\em et~al.}, }1996]{Haack-et-al1996}
{Haack}, H., {Scott}, E. R.~D., \& {Rasmussen}, K.~L. 1996.
\newblock {Thermal and shock history of mesosiderites and their large parent
  asteroid}.
\newblock {\em \gca}, {\bf 60}(14), 2609--2619.

\bibitem[\protect\citename{{Haba} {\em et~al.}, }2019]{Haba-et-al2019}
{Haba}, M.~K., {Wotzlaw}, J.-F., {Lai}, Y.-J., {Yamaguchi}, A., \&
  {Sch{\"o}nb{\"a}chler}, M. 2019.
\newblock {Mesosiderite formation on asteroid 4 Vesta by a hit-and-run
  collision}.
\newblock {\em \natg}, {\bf 12}(7), 510--515.

\bibitem[\protect\citename{{Hassanzadeh} {\em et~al.},
  }1990]{Hassanzadeh-et-al1990}
{Hassanzadeh}, J., {Rubin}, A.~E., \& {Wasson}, J.~T. 1990.
\newblock {Compositions of large metal nodules in mesosiderites: Links to iron
  meteorite group IIIAB and the origin of mesosiderite subgroups}.
\newblock {\em \gca}, {\bf 54}(11), 3197--3208.

\bibitem[\protect\citename{{Hopfe} \& {Goldstein},
  }2001]{Hopfe-and-Goldstein2001}
{Hopfe}, W.~D., \& {Goldstein}, J.~I. 2001.
\newblock {The metallographic cooling rate method revised: Application to iron
  meteorites and mesosiderites}.
\newblock {\em \mps}, {\bf 36}(1), 135--154.

\bibitem[\protect\citename{{Hsu}, }2003]{Hsu2003}
{Hsu}, W. 2003.
\newblock {Minor element zoning and trace element geochemistry of pallasites}.
\newblock {\em \mps}, {\bf 38}(8), 1217--1241.

\bibitem[\protect\citename{{Hubber} {\em et~al.}, }2013]{Hubber-et-al2013}
{Hubber}, D.~A., {Allison}, R.~J., {Smith}, R., \& {Goodwin}, S.~P. 2013.
\newblock {A hybrid SPH/N-body method for star cluster simulations}.
\newblock {\em \mnras}, {\bf 430}(3), 1599--1616.

\bibitem[\protect\citename{{Inutsuka}, }2002]{Inutsuka2002}
{Inutsuka}, S. 2002.
\newblock {Reformulation of Smoothed Particle Hydrodynamics with Riemann
  Solver}.
\newblock {\em \jcp}, {\bf 179}(1), 238--267.

\bibitem[\protect\citename{Iwasawa {\em et~al.}, }2015]{Iwasawa-et-al2015}
Iwasawa, M., Tanikawa, A., Hosono, N., Nitadori, K., Muranushi, T., \& Makino,
  J. 2015.
\newblock FDPS: A Novel Framework for Developing High-performance Particle
  Simulation Codes for Distributed-memory Systems.
\newblock {\em Pages  1:1--1:10 of:} {\em Proceedings of the 5th International
  Workshop on Domain-Specific Languages and High-Level Frameworks for High
  Performance Computing}.
\newblock WOLFHPC '15.
\newblock New York, NY, USA: ACM.

\bibitem[\protect\citename{{Iwasawa} {\em et~al.}, }2016]{Iwasawa-et-al2016}
{Iwasawa}, M., {Tanikawa}, A., {Hosono}, N., {Nitadori}, K., {Muranushi}, T.,
  \& {Makino}, J. 2016.
\newblock {Implementation and performance of FDPS: a framework for developing
  parallel particle simulation codes}.
\newblock {\em \pasj}, {\bf 68}(4), 54.

\bibitem[\protect\citename{{Lucy}, }1977]{Lucy1977}
{Lucy}, L.~B. 1977.
\newblock {A numerical approach to the testing of the fission hypothesis}.
\newblock {\em \aj}, {\bf 82}(Dec.), 1013--1024.

\bibitem[\protect\citename{{Mandler} \& {Elkins-Tanton},
  }2013]{Mandler-and-ElkinsTanton2013}
{Mandler}, B.~E., \& {Elkins-Tanton}, L.~T. 2013.
\newblock {The origin of eucrites, diogenites, and olivine diogenites: Magma
  ocean crystallization and shallow magma chamber processes on Vesta}.
\newblock {\em \mps}, {\bf 48}(11), 2333--2349.

\bibitem[\protect\citename{{Mason} \& {Jarosewich},
  }1973]{Mason-and-Jarosewich1973}
{Mason}, B., \& {Jarosewich}, E. 1973.
\newblock {The Barea, Dyarrl Island, and Emery meteorites and a view of the
  mesosiderites.}
\newblock {\em Mineralogical Magazine}, {\bf 39}(302), 204--215.

\bibitem[\protect\citename{{McCord} {\em et~al.}, }1970]{McCord-et-al1970}
{McCord}, T.~B., {Adams}, J.~B., \& {Johnson}, T.~V. 1970.
\newblock {Asteroid Vesta: Spectral Reflectivity and Compositional
  Implications}.
\newblock {\em \science}, {\bf 168}(3938), 1445--1447.

\bibitem[\protect\citename{{McSween} {\em et~al.}, }2011]{McSween-et-al2011}
{McSween}, H.~Y., {Mittlefehldt}, D.~W., {Beck}, A.~W., {Mayne}, R.~G., \&
  {McCoy}, T.~J. 2011.
\newblock {HED Meteorites and Their Relationship to the Geology of Vesta and
  the Dawn Mission}.
\newblock {\em \ssr}, {\bf 163}(1-4), 141--174.

\bibitem[\protect\citename{{Mittlefehldt} {\em et~al.},
  }1979]{Mittlefehldt-et-al1979}
{Mittlefehldt}, D.~W., {Chou}, C.~L., \& {Wasson}, J.~T. 1979.
\newblock {Mesosiderites and howardites: igneous formation and possible genetic
  relationships}.
\newblock {\em \gca}, {\bf 43}(5), 673,681--679,688.

\bibitem[\protect\citename{{Miyamoto} \& {Takeda},
  }1993]{Miyamoto-and-Takeda1993}
{Miyamoto}, M., \& {Takeda}, H. 1993.
\newblock {Rapid Cooling of Pallasite: Comparison of Chemical Zoning with
  Primitive Achondrites}.
\newblock {\em Meteoritics}, {\bf 28}(3), 404.

\bibitem[\protect\citename{{Monaghan}, }1992]{Monaghan1992}
{Monaghan}, J.~J. 1992.
\newblock {Smoothed particle hydrodynamics.}
\newblock {\em \araa}, {\bf 30}(Jan.), 543--574.

\bibitem[\protect\citename{{Nathues} {\em et~al.}, }2015]{Nathues-et-al2015}
{Nathues}, A., {Hoffmann}, M., {Sch{\"a}fer}, M., {Thangjam}, G., {Le Corre},
  L., {Reddy}, V., {Christensen}, U., {Mengel}, K., {Sierks}, H., {Vincent},
  J.-B., {Cloutis}, E.~A., {Russell}, C.~T., {Sch{\"a}fer}, T.,
  {Gutierrez-Marques}, P., {Hall}, I., {Ripken}, J., \& {B{\"u}ttner}, I. 2015.
\newblock {Exogenic olivine on Vesta from Dawn Framing Camera color data}.
\newblock {\em \icarus}, {\bf 258}(Sep), 467--482.

\bibitem[\protect\citename{{O'Brien} \& {Sykes}, }2011]{OBrien-and-Sykes2011}
{O'Brien}, David~P., \& {Sykes}, Mark~V. 2011.
\newblock {The Origin and Evolution of the Asteroid
  Belt{\textemdash}Implications for Vesta and Ceres}.
\newblock {\em \ssr}, {\bf 163}(1-4), 41--61.

\bibitem[\protect\citename{{Peplowski} {\em et~al.},
  }2013]{Peplowski-et-al2013}
{Peplowski}, P.~N., {Lawrence}, D.~J., {Prettyman}, T.~H., {Yamashita}, N.,
  {Bazell}, D., {Feldman}, W.~C., {Le Corre}, L., {McCoy}, T.~J., {Reddy}, V.,
  {Reedy}, R.~C., {Russell}, C.~T., \& {Toplis}, M.~J. 2013.
\newblock {Compositional variability on the surface of 4 Vesta revealed through
  GRaND measurements of high-energy gamma rays}.
\newblock {\em \mps}, {\bf 48}(11), 2252--2270.

\bibitem[\protect\citename{{Prinz} {\em et~al.}, }1980]{Prinz-et-al1980}
{Prinz}, M., {Nehru}, C.~E., {Delaney}, J.~S., {Harlow}, G.~E., \& {Bedell},
  R.~L. 1980 (Mar).
\newblock {ALHA 77219: a New Antarctic Mesosiderite and a Comparison with Other
  Mesosiderites and Related Achondrites}.
\newblock {\em Pages  899--901 of:} {\em Lunar and Planetary Science
  Conference}.
\newblock Lunar and Planetary Science Conference.

\bibitem[\protect\citename{{Rubin} \& {Mittlefehldt},
  }1993]{Rubin-and-Mittlefehldt1993}
{Rubin}, A.~E., \& {Mittlefehldt}, D.~W. 1993.
\newblock {Evolutionary History of the Mesosiderite Asteroid: A Chronologic and
  Petrologic Synthesis}.
\newblock {\em \icarus}, {\bf 101}(2), 201--212.

\bibitem[\protect\citename{{Ruzicka} {\em et~al.}, }1997]{Ruzicka-et-al1997}
{Ruzicka}, A., {Snyder}, G.~A., \& {Taylor}, L.~A. 1997.
\newblock {Vesta as the HED Parent Body: Implications for the Size of a Core
  and for Large-Scale Differentiation}.
\newblock {\em \mps}, {\bf 32}(6), 825--840.

\bibitem[\protect\citename{Scott, }1977]{Scott1977}
Scott, E. R.~D. 1977.
\newblock Geochemical relationships between some pallasites and iron
  meteorites.
\newblock {\em Mineralogical Magazine}, {\bf 41}(318), 265–272.

\bibitem[\protect\citename{{Scott} {\em et~al.}, }2001]{Scott-et-al2001}
{Scott}, E. R.~D., {Haack}, H., \& {Love}, S.~G. 2001.
\newblock {Formation of mesosiderites by fragmentation and reaccretion of a
  large differentiated asteroid}.
\newblock {\em \mps}, {\bf 36}(6), 869--891.

\bibitem[\protect\citename{{Sugiura}, }2020]{Sugiura2020}
{Sugiura}, K. 2020.
\newblock {\em Development of a Numerical Simulation Method for Rocky Body
  Impacts and Theoretical Analysis of Asteroidal Shapes}. 1 edn.
\newblock Springer, Singapore.

\bibitem[\protect\citename{{Tillotson}, }1962]{Tillotson1962}
{Tillotson}, J.~H. 1962 (Jul).
\newblock {\em {Metallic Equations of State For Hypervelocity Impact}}.
\newblock Tech. rept.

\bibitem[\protect\citename{{Trinquier} {\em et~al.},
  }2007]{Trinquier-et-al2007}
{Trinquier}, A., {Birck}, J.-L., \& {All{\`e}gre}, C.~J. 2007.
\newblock {Widespread $^{54}$Cr Heterogeneity in the Inner Solar System}.
\newblock {\em \apj}, {\bf 655}(2), 1179--1185.

\bibitem[\protect\citename{{Yamaguchi} {\em et~al.},
  }2011]{Yamaguchi-et-al2011}
{Yamaguchi}, A., {Barrat}, J.-A., {Ito}, M., \& {Bohn}, M. 2011.
\newblock {Posteucritic magmatism on Vesta: Evidence from the petrology and
  thermal history of diogenites}.
\newblock {\em \jgr}, {\bf 116}(E8), E08009.

\end{thebibliography}

\end{document}